\documentclass{iopart}
\usepackage{graphicx} 
\usepackage{amssymb}
\usepackage{bm}
\usepackage{units}
\usepackage{float}
\usepackage{subfigure}
\usepackage{url}
\usepackage{xcolor}
\usepackage[margin=1.5in]{geometry}

\newcommand{\td}[1]{{#1}}  

\newcommand{\dl}[1]{{#1}}

\newcommand{\pdet}{P_\mathrm{det}}
\newcommand{\emax}{\varepsilon_{\max}}
\newcommand{\dmid}{d_\mathrm{mid}}

\begin{document}

\title[LIGO-Virgo compact binary merger selection function]{A physically modelled selection function for compact binary mergers in the LIGO-Virgo O3 run and beyond}
\author{Ana Lorenzo-Medina$^1$, Thomas Dent$^1$}
\address{$^1$ Instituto Galego de F\'{i}sica de Altas Enerx\'{i}as, Universidade de Santiago de Compostela, 15782 Santiago de Compostela, Galicia, Spain}
\ead{analorenzo.medina@usc.es,thomas.dent@usc.es}
       

\begin{abstract}
    Despite the observation of nearly 100 compact binary coalescence (CBC) events up to the end of the Advanced gravitational-wave (GW) detectors' third observing run (O3), there remain fundamental open questions regarding their astrophysical formation mechanisms and environments. Population analysis should yield insights into these questions, but requires careful control of uncertainties and biases. GW observations have a strong selection bias: this is due first to the dependence of the signal amplitude on the source's (intrinsic and extrinsic) parameters, and second to the complicated nature of detector noise and of current detection methods. 
    In this work, we introduce a new physically-motivated model of the sensitivity of GW searches for CBC events, aimed at enhancing the accuracy and efficiency of population reconstructions. 
    In contrast to current methods which rely on re-weighting simulated signals (injections) via importance sampling, we model the probability of detection of binary black hole (BBH) mergers as a smooth, analytic function of source masses, orbit-aligned spins, and distance, 
    fitted to accurately match 
    injection results. 
    The estimate can thus be used for population models whose signal distribution over parameter space differs significantly from the injection distribution. 
    Our method has already been used in population studies such as reconstructing the BBH merger rate dependence on redshift. 
\end{abstract}

\maketitle

\section{Introduction}

Gravitational wave (GW) astronomy has made a number of breakthroughs in recent years, with the first detection of merging black holes 
by the interferometric gravitational-wave detector network on 14 September 2015~\cite{firstobv}.  Coalescences of compact binaries (CBC), such as black hole or neutron star binaries, are a particularly interesting source for broad-band detectors, since the signal from an inspiral is quite strong compared to other types of GWs, and the binary system properties can be measured with high precision for some parameters of astrophysical interest, due to the large number of signal cycles observable in the sensitive frequency band~\cite{Cutler:1994ys}.  \td{With the current catalog of close to 100 observed binaries~\cite{KAGRA:2021vkt} many unanswered questions about such sources remain, such as the origin of GW-emitting black hole binary (BBH) systems, it being debatable whether they have properties different from~\cite{Fishbach:2021xqi} or similar to~\cite{Belczynski:2021agb} black holes detected via X-ray emission}. 

Many formation channels and mechanisms have been proposed for the BBH observed via GW emission (see~\cite{Spera:2022byb} for an overview).  Among the most intensively studied are isolated binary evolution (e.g.\ \cite{Belczynski:2017gds,Stevenson:2017tfq}, and dynamical binary formation in globular or other stellar clusters (e.g.\ \cite{Rodriguez:2016kxx,OLeary:2016ayz}) or around galactic centres~(e.g.\ \cite{Mckernan:2017ssq}).  Each of these channels may have distinctive signatures in the distribution of merging binaries over masses, spins and redshift, though subject to many theoretical and modelling uncertainties (e.g.~\cite{Belczynski:2021zaz}; see \cite{pop2,o3b_pop,LIGOScientific:2016vpg} for more detailed discussion). 
Currently, no single formation channel appears capable of explaining the entire population~\cite{Zevin:2020gbd}, implying that a more complex mixture of channels will eventually be crucial to properly describe observations. 
\td{The current detector network is only sensitive over a limited range of redshift: thus, although some studies find hints of population evolution over cosmic history, such findings are so far strongly dependent on the choice of methods~\cite{Fishbach:2021yvy,Tiwari:2021yvr,Karathanasis:2022rtr,Biscoveanu:2022qac,Rinaldi:2023bbd}}. 
Since various formation channels are expected to have different dependencies on redshift~\cite{Mapelli:2021gyv,vanSon:2021zpk,Belczynski:2022wky}, tracking such evolution will become increasingly important to untangle them as detector sensitivity increases (e.g.~\cite{Ng:2020qpk}). 

The set of binary mergers that we observe comes with a significant bias as compared to the actual astrophysical BBH population, because of selection effects, which are crucial to take into account in population studies \td{(see e.g.~\cite{rate_predictions} for an early review)}. 
The most direct selection effect, and the one currently limiting our ability to probe cosmic evolution, is that nearby (low redshift) sources are more likely to be detected than far (high redshift) ones.  Nearly as important is the dependence of signal amplitude on binary masses: for the inspiral phase of the signal, this varies as (a power of) the chirp mass $\mathcal{M} = (m_1m_2)^{3/5} / (m_1+m_2)^{1/5}$, while for the later merger-ringdown phase the binary total mass has the most influence.  The spins of binary components also affect the GW signal detectability \cite{chieff,chieff_effects}, primarily via a longer or shorter signal duration with the merger at higher (lower) frequency for spins aligned (anti-aligned) with the orbit.  Spins in the plane of the orbit, which give rise to precession effects, also affect detection, primarily since the existing searches use templates with only orbit-aligned component spins, which therefore cannot fully match signals with precession (e.g.~\cite{Harry:2016ijz}); similar considerations apply to non-dominant multipole (``higher mode'') emission which is not included in current template banks~\cite{Capano:2013raa,Harry:2017weg}.  \td{Signals from binaries with non-negligible residual orbital eccentricity will also be less efficiently recovered~\cite{Brown:2009ng,Gadre:2024ndy}; studies of eccentricity in combination with other effects will eventually be required, with accurate waveform modelling as a technical limiting factor (see e.g.\ discussions in~\cite{Romero-Shaw:2022fbf}).}

The interplay of search sensitivity with these astrophysical properties of binaries makes it crucial to have an accurate quantitative estimate of selection effects when carrying out population analysis.  This is provided by an estimate of the probability of detection $\pdet$, or selection function, of binary mergers: an accurate estimate enables measurement of the rates of binary merger events, and reconstruction of their population distribution as a function of component masses, spins and redshift~\cite{pop2,o3b_pop}. 

Events are considered to be detected on the basis of their statistical significance assigned by search pipelines, i.e.\ matched-filtering algorithms looking for signals in the data~\cite{Allen:2005fk,Usman:2015kfa,Messick:2016aqy,Adams:2015ulm,Luan:2011qx}.  Each candidate event has a ranking statistic value calculated as a predefined function of the event parameters.  The significance may be quantified by the false alarm rate (FAR), which is the expected rate of candidates in the pipeline resulting from detector noise which are ranked equal to or higher than the candidate under consideration. 
A set of candidates of well defined purity, in terms of the expected number of noise events, may be obtained by imposing a pre-defined maximum threshold on FAR.
BBH candidates have typically been considered for population analysis if their FAR is below a threshold $1/\mathrm{yr}$; in the case of several pipelines, \td{one may apply the threshold to the most significant result between pipelines for a given candidate~\cite{pop2,o3b_pop}.\footnote{\td{Other methods of quantifying event significance by pooling multiple pipeline results may be considered~\cite{Banagiri:2023ztt,Lukina_Combined}.}}}

The detection fraction of a population described by a model PDF $\pi(\theta|\Lambda)$ with (hyper)parameters $\Lambda$ is a crucial element in population inference (e.g.~\cite{Mandel:2018mve,importance_sampling}): it is defined by
\begin{equation}
    \xi (\Lambda) = \int \pdet (\theta) \pi (\theta | \Lambda) d(\theta) \, ,
\end{equation}
where $\pdet (\theta)$ stands for the probability that an event with parameters $\theta$ would be detected (found) by a particular search or a combination of searches. For binary population analyses, $\pdet$ is estimated \td{indirectly} by simulating CBC signals from a reference BBH population, adding them to real detector data and analysing the ``injected'' data with search pipelines to record which ones are detected~\cite{o3b_pop}. 
For the detection fraction to be well defined, we must restrict the time of arrival (considered as an event parameter) to periods where any detectors were observing, and effectively calculate average detectability over such periods.

Since the previous integral cannot be evaluated analytically, importance sampling may be used instead to estimate it using a weighted Monte Carlo integral \td{(see e.g.~\cite{NumRecipes})} over found injections, i.e.\ those satisfying the chosen search significance threshold:
\begin{equation}
    \hat{\xi}(\Lambda) = \frac{1}{N_\mathrm{inj}} \sum^{N_\mathrm{det}}_{j=1} \frac{\pi(\theta_j | \Lambda)}{p_\mathrm{draw} (\theta_j)} \equiv 
    \frac{1}{N_\mathrm{inj}} \sum^{N_\mathrm{det}}_{j=1} w_j
    \, ,
\end{equation}
in which $N_\mathrm{inj}$ is the total number of injections, $j = 1,\ldots,N_\mathrm{det}$ labels the found (detected) injections, and $p_\mathrm{draw}$ is a fiducial distribution from which the samples are drawn~\cite{Tiwari:2017ndi,importance_sampling}. 

Importance sampling is a convenient approach since it only requires one pre-determined injection analysis campaign to be performed by searches, which is then used to estimate selection effects for any population models considered.  
However, this procedure comes with uncertainty in the estimation of the selection integral: the precision of the estimation depends on how well the proposal model $\pi$ matches the form of $p_\mathrm{draw}$. There must be enough samples drawn and detected by search pipelines so that this uncertainty does not alter the shape of the population model posterior~\cite{importance_sampling,Essick:2022ojx,Talbot:2023pex}. 
The precision may be quantified via the effective number of independent draws (sample size), $N_\mathrm{eff}$, that contribute to estimate $\xi$ for a given model.  This effective sample size is defined as the square of the estimated fraction $\hat{\xi}$ divided by its variance due to Monte Carlo uncertainty \cite{Talbot:2023pex}; in terms of the weights $w_j = \pi(\theta_j) / p_\mathrm{draw} (\theta_j)$, $N_\mathrm{eff}$ varies as
\begin{equation}
 \frac{\hat{\xi}^2}{\mathrm{Var(\hat{\xi})}} \equiv N_\mathrm{eff} \simeq \frac{(\sum_j w_j)^2}{\sum_j w_j^2 } \, .
\end{equation}
%
Then depending on the model $\pi(\theta|\Lambda)$, for a fixed injection set $N_\mathrm{eff}$ may or may not be high enough to obtain an accurate population inference; as a general guide \td{$N_\mathrm{eff} > 4N_\mathrm{obs}$ has been required (where $N_\mathrm{obs}$ is the number of events in the population catalog)}  ~\cite{importance_sampling,Essick:2022ojx}.  
Then, if we discount the possibility of requiring more injections to be analyzed ad-hoc, in general at high computational cost, a large class of models cannot be studied using a generic (broadly/uniformly distributed) injection set~\cite{Talbot:2023pex}.  For example, population models where $\pi(\theta|\Lambda)$ is strongly peaked around a specific point, if the density of injections $p_\mathrm{draw}(\theta)$ is featureless (e.g.~\cite{Galaudage:2021rkt}); or in general, where some $w_i$ are $\gg 1$ (see also~\cite{chatterjee_importance}). 

As an alternative, using an analytical function -- or in general, a continuous function -- that accurately describes the probability of detection of BBH signals will make it possible to estimate $\pdet$ without such restrictions on the model and original injection sampling distributions. \td{As an initial effort towards such a description}, 
the dependence on masses and spins of the selection function can be modelled via ``semi-analytic'' methods 
with various possible corrections.  The interferometer network sensitivity is approximated by evaluating the expected (optimal) signal-to-noise ratios (SNRs) of signals for fixed, ideal detector power spectral densities (PSDs).  The probability of detection is then obtained by assuming that the network is sensitive to CBCs with SNR above a fiducial threshold value~\cite{rate_predictions,Fishbach:2017zga}, which may be tuned to approximate the number of actual signals, or injections detected by search pipelines~\cite{Essick:2023toz,Gerosa:2024isl}. 

However, a simple optimal SNR threshold is not necessarily an accurate approximation for $\pdet$, due to many factors which violate the assumptions behind the ``semi-analytic'' calculation.  Beyond the effect of the noise realization on matched filter SNR \td{(e.g.~\cite{Gerosa:2024isl})}, for instance, the number of detectors observing and their sensitivity varies over time\td{~\cite{pycbc, Sachdev:2019vvd, mbta, Davies:2022thw}};  
detector noise is also not completely described by a single set of PSDs, being not perfectly Gaussian and stationary.  In reality there are excess noise transients (``glitches'', see e.g.~\cite{LIGO:2021ppb,Virgo:2022ysc}) that may be partly suppressed by search pipelines employing additional quantities beyond the SNR, and other non-ideal noise behaviour\td{~\cite{Zackay:2019kkv,Mozzon:2020gwa}}.  \td{In general the search ranking statistic may have additional dependences beyond the SNR, for example explicit dependence on masses and/or spins~\cite{gstlal,Kumar:2024bfe}\footnote{\td{Such dependence amounts to modelling a signal population for which the search is optimized~\cite{Dent:2013cva}: like other aspects of the event ranking, this must be determined in advance of performing the search.}}, and/or on detector outputs other than strain which are correlated with transient noise~\cite{Essick:2020qpo,Godwin:2020weu,Davis:2022cmw}. }

Hence, corrections to the semi-analytic approach will in general be necessary.  A simple grid-based approach to obtain such corrections by fitting the mass- and (aligned) spin-dependent sensitive volume, i.e.\ the integral of $\pdet$ over space properly weighted by comoving volume-time, to search injection results was developed in~\cite{pdetbefore} and applied in~\cite{pop2,o3b_pop}.  
Machine learning techniques using density estimation on found injections to get a continuous and generative model for $\pdet$ were demonstrated in~\cite{pdetbefore_2}; however, neither method was applied to obtain the distance or redshift dependence of $\pdet$, which is crucial for cosmological investigations such as~\cite{Rinaldi:2023bbd}.  One obstacle to directly estimating $\pdet(d_L)$ is the sparsity of injections at small distances, if distributed with (comoving) volume. 

The strategy we pursue here to overcome this limitation is to fit a continuous and physically motivated $P_\mathrm{det}$ function over distance, masses and, \td{ultimately}, spins.  While the basic functional form is derived from the known theoretical dependence of SNR on distance and chirp mass, we introduce corrections to account for non-ideal detector behaviour, and for higher order effects of masses and spins on an event's detectability in actual search pipelines.  The fit is adjusted to maximize the likelihood of found injections in the O3 sensitivity data release corresponding to GWTC-3~\cite{o3,KAGRA:2021vkt}.  

The remainder of the paper is structured as follows: in Section \ref{sec:methods} we lay out the methods used in this work, as well as the functions derived to fit $P_\mathrm{det}$ accurately over masses and distance. In Sec.~\ref{sec:results} we present the results obtained from the fit to the O3 injection set, and validate the first by comparison of CDFs and Kolmogorov-Smirnov (KS) tests. In Sec.~\ref{sec:spins} we add a dependence to $P_\mathrm{det}$ on the effective spin, and demonstrate a resulting improvement in the fit accuracy.  In Sec.~\ref{sec:uncertainties} we quantify the statistical uncertainties in our fitted function for $P_\mathrm{det}$ via bootstrap resampling.  Finally, in Sec.~\ref{sec:discussion} we summarize the work and discuss possible implications and further developments. 

\section{Methods}\label{sec:methods}

The aim of this project is to obtain accurate estimates of the selection function, or probability of detection, for gravitational-wave sources considering the only type of system so far detected, compact binaries of neutron stars or black holes. The selection function depends strongly on the properties of the binary components as well as the source's distance from Earth, and thus on the redshift. 

LIGO-Virgo-KAGRA (LVK) O3 BBH search sensitivity estimates \cite{o3} were used for the analysis. This data release contains sets of simulated signals (injections) from binary black holes that can be used to estimate the search sensitivity corresponding to the GWTC-3 catalog~\cite{gwtc3}. 

Regarding the analysis presented in this work we are considering the binary source component masses and redshift, hence the parameters determining the detectability of signals will be the redshifted masses $m_{1z}$, $m_{2z}$ (masses in the detector's frame) and the luminosity distance $d_L(z)$.  Selection effects are thus described via a function $\pdet(m_{1z}, m_{2z}, d_L)$ giving the probability of detection for a merger occurring during an observing period. The probability is understood to be marginalized over extrinsic (location and orientation) parameters, over time of arrival and over detector noise realizations, and also, in sections~\ref{sec:methods} and~\ref{sec:results}, over binary component spins.  

\subsection{Sigmoid distance dependence model}

\begin{figure}[tbp]
	\centering
	\subfigure[{$\pdet(d_L)$ for fixed sky position and orientation}]{\includegraphics[width=0.46\textwidth]{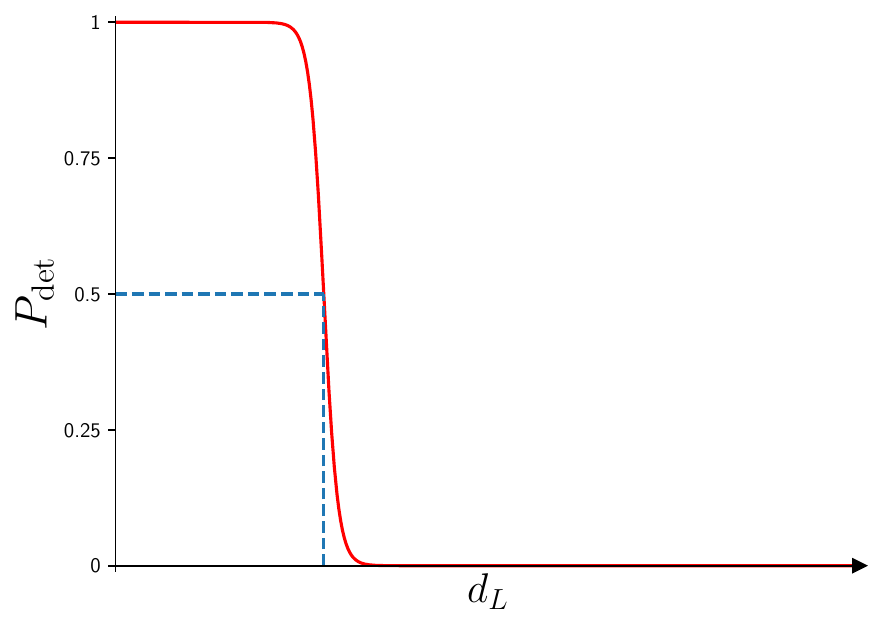}}
	\hspace{0.4cm}
	\subfigure[$\pdet(d_L)$ marginalizing over sky position and orientation]{ \includegraphics[width=0.46\textwidth]{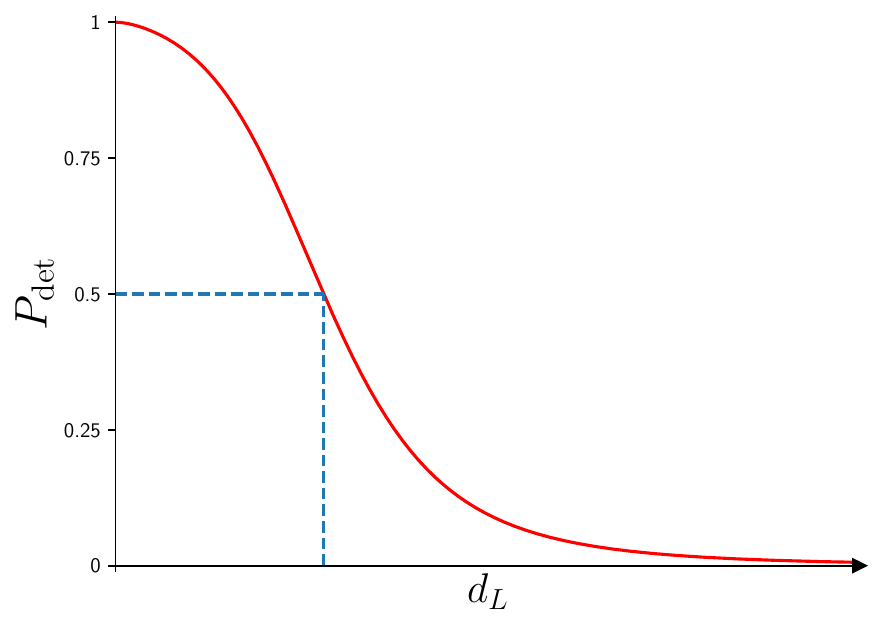}}
	\caption{Schematic dependence of search sensitivity on luminosity distance for fixed binary intrinsic parameters.  The blue dashed lines indicate $\dmid$, the distance at which detection probability falls to $0.5$.}
	\label{fig:eps}
\end{figure}

Considering a source of fixed intrinsic parameters (binary component masses and spins), the dependence on distance of search sensitivity is mainly determined by \td{the sky position and orientation of the source relative to a detector network (assumed to have stationary sensitivity), described by various angular parameters}.  If the angular parameters were also fixed, then the amplitude and expected SNR are simply inversely proportional to distance: then, assuming the source is detected if SNR is above a given threshold, $\pdet(d_L)$ would be almost 1 for some range of distances up to a given point, and then close to 0 for larger distances, as represented in Fig.~\ref{fig:eps}a. 
The detection probability has the shape of a sigmoid instead of a step function because of noise in the detector: in an actual search, detection will be determined by the matched filter SNR, a function of the data which contains a stochastic term resulting from the detector noise realization in addition to the expected SNR. 

Since the sources' positions and orientations are \textit{a priori} unknown, and isotropically distributed throughout the sky, we need to consider all the possible values of the angular parameters and average over them. As a result of this marginalization, the functional form of the search sensitivity takes a form more similar to Fig.~\ref{fig:eps}b. 
These effects on search sensitivity were quantified by Finn \& Chernoff (FC)~\cite{FC}, predicting the form of the probability of detection function over distance by considering geometrical factors in the expected SNR for the case of only one detector with fixed sensitivity. 

Although the true functional dependence of $\pdet(d_L)$ in a realistic case will have many corrections relative to this idealized calculation, our fundamental model assumption is that such corrections are small.  Thus, as a starting point we use a parameterized sigmoid function that is able to fit the FC data (\cite{FC} Table I) faithfully, while also allowing for minor deviations.  The shape of the sigmoid is controlled by four parameters $\emax, \alpha, \delta$ and $\gamma$, out of which $\alpha = 2.05$ is fixed to the value obtained in a direct fit to FC data and the other 3 will be free model parameters. The probability of detection, for a binary with given intrinsic parameters, thus reads as
\begin{equation}
 \pdet = \frac{\emax}
  {1 + {\hat{d}_L}^\alpha \exp \left[\gamma(\hat{d}_L - 1) + \delta(\hat{d}_L^2 - 1) \right]},
  \label{pdet}
\end{equation} 
where $\hat{d}_L \equiv d_L / \dmid$ is the luminosity distance, scaled by the ``midpoint distance'' $\dmid$ at which the sigmoid falls to 50\% of its maximum value, and $\emax$ gives the detection probability in the limit $d_L \rightarrow 0$. 
Specifically, within a given observing run, the adjustable parameters are intended to model the effects (averaged over the run) of both stationary Gaussian detector noise, and non-stationary behaviour affecting detector sensitivities, including the observation duty cycles of different detectors in the network. 

Furthermore, we allow the search sensitivity in the limit of very loud signals $\emax$ to be less than 1, which may happen due to effects of precession, higher modes, etc., causing the GW signal to differ significantly from the templates used in searches (e.g.~\cite{spin_pipeline,Harry:2016ijz,Capano:2013raa}); for published LVK results, such templates include only the dominant multipole emission from non-precessing binaries, as advocated in~\cite{chieff}. 

Having modelled the dependence on amplitude and distance for fixed intrinsic binary parameters, we then add a dependence on these intrinsic parameters by writing $\dmid (m_{1z}, m_{2z})$ and $\emax(m_{1z}, m_{2z})$ as a function of the redshifted masses (we will also allow for component spin dependence in Sec.~\ref{sec:spins}).  In the low mass limit, $\dmid$ \td{is expected to take the form} $\dmid \sim D_0 f_d \mathcal{M}_z^{5/6}$, where $D_0$ \td{is a constant corresponding to average detector sensitivity over a given observing run}; i.e.\ sensitive distance should scale with $\mathcal{M}_z^{5/6}$ for the simplest inspiral model, where the redshifted chirp mass is given by $\mathcal{M}_z = \eta^{3/5}M_z$ in terms of the redshifted binary total mass $M_z = m_{1z} + m_{2z}$ and the symmetric mass ratio $\eta = m_1 m_2/M^2 \equiv m_{1z}m_{2z}/M_z^2$.  The remaining dependence on masses, corresponding to corrections to this lowest order inspiral SNR power law behaviour, is then to be fitted by a function $f_d (M_z, \eta)$. 
Taking into account that $\emax$ should not be higher than 1 and always positive, we take a functional form $\emax \sim 1 - \exp (f_\varepsilon)$, with $f_\varepsilon$ a function of intrinsic parameters.

At this point, the functions providing mass dependence to $\dmid$ and $\emax$ could be any arbitrary combination of $M_z$ and $\eta$. In order to gain more insight into the needed corrections, we perform mass binned fits to the detected injections released with the GWTC-3 catalog.

\subsection{Mass binned fits}
\label{ss:massbin}

In order to gain insight into the functional forms of 
$f_d$ and $f_\varepsilon$ that will be able to fit injection results over the BBH mass space, we first fit our analytical model of $\pdet$ as a function of distance for fixed masses, or rather in mass bins, by numerically maximizing the likelihood of the injection campaign search results.\footnote{Code to evaluate our $\pdet$ model and fit its parameters to injection results, and the resulting maximum likelihood parameter values, are publicly available at \url{https://github.com/AnaLorenzoMedina/cbc_pdet .}}  
In this binned fit procedure, rather than being continuous functions, $\dmid$ and $\emax$ are treated as scalar parameters, fitted separately for each mass bin; \td{we choose bins equally spaced in the logarithms of $m_1$ and $m_2$, as they then contain comparable counts of found injections with FAR $\leq 1$/yr~\cite{o3b_pop}, at least for mass ratios close to unity}. 
For a given mass bin labeled as $\cdot^M$ we use a inhomogeneous Poisson likelihood for the found injections: 
\begin{equation}
    \ln \mathcal{L}^M = -\mu^M + \sum_i \ln \lambda^M (d_L^i) \, ,
\end{equation}
where $\lambda^M$ stands for the expected density of found injections at a given distance,
\begin{equation}
    \lambda^M(d_L) = \pdet(d_L) \, p(d_L) \, N_\mathrm{tot}^{M} \, ,
\end{equation}
where $\pdet(d_L)$ is evaluated using given values of $\dmid$ and $\emax$; we write $N_\mathrm{tot}^{M} = $ $ N_\mathrm{total}$ $\int\!\!\int \! p(m_1,m_2) \, dm_1 dm_2$ for the expected total number of injections in the bin, where the distributions $p(\theta)$ are the sampling probability density functions (pdf) for $\vec{\theta}=\{m_1, m_2, d_L\}$ (described in \ref{appendix:pdf} for O3 injections), $N_\mathrm{total}$ is the total number of injections originally generated, and $\mu$ stands for the expected total number of found injections in the bin:
\begin{equation}
    \mu^M \equiv N_{\exp} = \int _0 ^{\infty} \lambda^M(d_L) \, \mathrm{d}d_L \, .
\end{equation}

Although $\dmid$ and $\emax$ vary between bins, we use the same shape parameter values $\delta$ and $\gamma$ over the whole mass range, since we assume $\pdet$ should have the same shape regardless of the mass. 
We then perform a fit in two alternating stages.  In the `sigmoid fitting' stage, $\dmid$ and $\emax$ are held fixed for all bins and the sigmoid shape is fitted by maximizing the total log likelihood $\sum_M \ln \mathcal{L}^M$ over $\delta$ and $\gamma$.  In the `mass fitting' stage, $\delta$ and $\gamma$ are instead fixed and the values of $\dmid$ and $\emax$ are fitted separately for each bin by maximizing $\ln \mathcal{L}^M$.  The two stages are alternated until the total likelihood has converged to a global maximum \td{(see e.g.~\cite{10.1007/3-540-45631-7_39} for a general discussion of ``alternating optimization'')}. 
For optimization at each stage the Nelder-Mead simplex algorithm was used, a typical direct search method for multidimensional unconstrained minimization (e.g.~\cite{neldermead}). 
\begin{figure}[tbp]
	\centering 
	\includegraphics[width=0.9\linewidth]{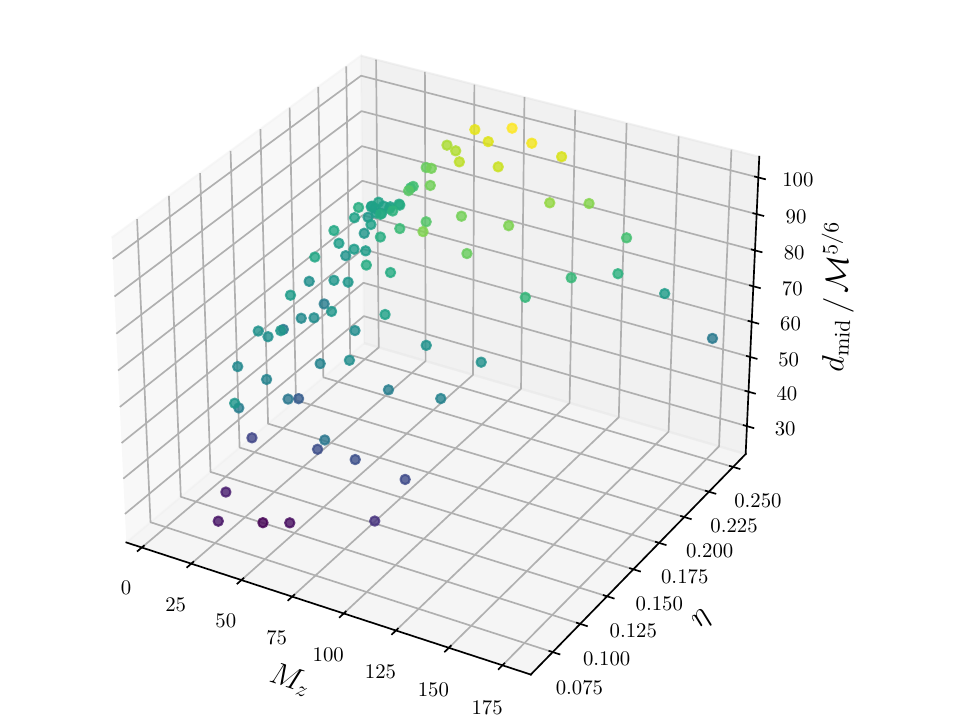}
	\caption{Optimized \td{sensitive distance} $\dmid$ estimates from the mass binned fits.  \td{Points are colored according to the z-axis values $\dmid / \mathcal{M}_z^{5/6}$ for clarity.}}
	\label{fig:dmid_values}
\end{figure}

Figure~\ref{fig:dmid_values} shows the resulting optimized values of $\dmid$, having scaled out a factor of $\mathcal{M}_z^{5/6}$.  While all the values are comparable in magnitude, we see strong linear and quadratic dependencies on $M_z$ at higher $\eta$, decreasing as the symmetric mass ratio gets smaller. 
Taking into account this behaviour, we first proposed the following expression to describe the global mass dependence:\footnote{\td{The $a$ coefficient subscripts, in order, represent powers of $M_z$ and $(1-4\eta)$.}}
\begin{equation}
\fl\;\; \dmid^*(m_{1z}, m_{2z}) = D_0 \mathcal{M}_z^{5/6} \left[1 + a_{10}M_z + a_{20}M_z^2 
  + (a_{01} + a_{11}M_z + a_{21}M_z^2) (1 - 4\eta) \right], 
\end{equation}
in which we are separating out the dependence away from equal masses (i.e.\ for $\eta < 1/4$) where there may be higher uncertainty, treating this as a correction for unequal mass binaries.  This first fit was already used in a population analysis to investigate the evolution of the BBH population over redshift~\cite{Rinaldi:2023bbd}. However this function 
has the disadvantage that it may become negative, which would be physically disallowed; therefore \td{for our main results we use a slightly modified form with corrections via an exponential}:
\begin{equation} \label{eq:dmid_fd}
 \dmid(m_{1z}, m_{2z}) = D_0 \mathcal{M}_z^{5/6} \exp (f_d) \, ,
\end{equation}
\begin{equation} \label{eq:fd_nospin}
 f_d =  a_{10}M_z + a_{20}M_z^2 
  + (a_{01} + a_{11}M_z + a_{21}M_z^2) (1 - 4\eta) \, .
\end{equation} 

\begin{figure}[tbp]
	\centering 
	\includegraphics[width=0.65\linewidth]{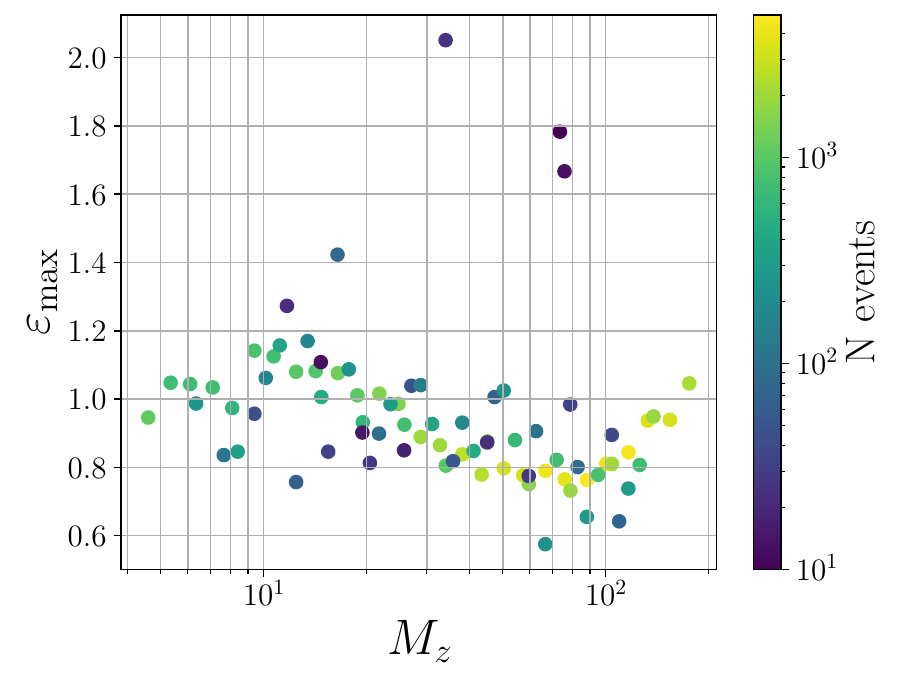}
	\caption{Maximum search efficiency $\emax$ fitted values for each mass bin: the color bar shows the number of detected events (injections) in each bin.}
	\label{fig:emax_values}
\end{figure}

\td{We next consider the functional dependence of $\emax$ (the detection probability in the limit of zero distance) by inspecting the optimized values of $\emax$ for binned fits} in Fig.~\ref{fig:emax_values}. Here, we allow (in principle unphysical) $\emax$ values greater than 1 which may be caused by small number statistics; although the total number of found injections in a bin may be large, due to their distribution over distance, the detection probability at small distance is typically badly constrained.
The \td{trend of variation is most apparent over} linear total mass, therefore given the restrictions stated previously we chose the following function: 
\begin{equation} \label{eq:emax_nospin}
    \emax (m_{1z}, m_{2z}) = 1 - \exp (b_0 + b_1 M_z + b_2 M_z^2) \, .
\end{equation}

\subsection{Global fits: mass dependence}

Given the functional forms of $f_d$ and $f_\varepsilon$ described in the previous section, we have everything we need to make a global fit, that is, obtain a smooth function of $\pdet$ in terms of both distance and masses. To optimize the parameters of the global function, the previous expressions for the likelihood need to be generalised:
\begin{equation}
    \ln \mathcal{L} = -\mu + \sum_i \ln \lambda (d_L^i, m_1^i , m_2^i) \, ,
    \label{global_likelihood}
\end{equation}
\begin{equation}
    \lambda(d_L,m_1 , m_2) = \pdet(d_L, m_1, m_2) \, p(d_L) \, p(m_1, m_2) \, N_{\mathrm{tot}} \, ,
    \label{smalllambda}
\end{equation}
\begin{equation}
    \mu \equiv N_{\exp} = \int\!\!\int\!\!\int\! \lambda(d_L, m_1 , m_2) \, \mathrm{d}m_1 \, \mathrm{d}m_2 \, \mathrm{d}d_L \, .
    \label{nexp}
\end{equation}
The triple integral in (\ref{nexp}) could in principle be computed numerically, but at high computational cost.  Instead, we obtain an accurate approximation from $N_\mathrm{exp} \simeq \sum_i \pdet(d_{Li}, m_{1i}, m_{2i})$, evaluating over all injections (both found and missed) as these are already correctly distributed as $p(d_L)p(m_1,m_2)$.

The sigmoid parameters $\gamma$ and $\delta$, as well as the new parameters from $\dmid$ and $\emax$, were fitted to the GWTC-3 injections by maximizing the likelihood (\ref{global_likelihood}).  Similarly to the binned fit, we perform optimization in two alternating steps. Firstly, the likelihood is maximized optimizing over the $\dmid$ parameters, $\{D_0, a_{10}, a_{20}, a_{01}, a_{11}, a_{21} \}$, and the rest of them are fixed. In step two the optimization takes place over the parameters  controlling the shape of the sigmoid curve, namely $\{\gamma, \delta, b_0, b_1, b_2 \}$, using the optimized $\dmid$ values gotten in step 1.  Then the two steps are repeated until the difference in the log likelihood is less than $0.01$, in order to ensure a stable maximum value.

\section{Results and discussion}\label{sec:results}

\begin{figure}[tbp]
	\centering
	\subfigure[$\dmid$ in terms of $m_{1z}$ and $m_{2z}$.]{\includegraphics[width=0.49\textwidth]{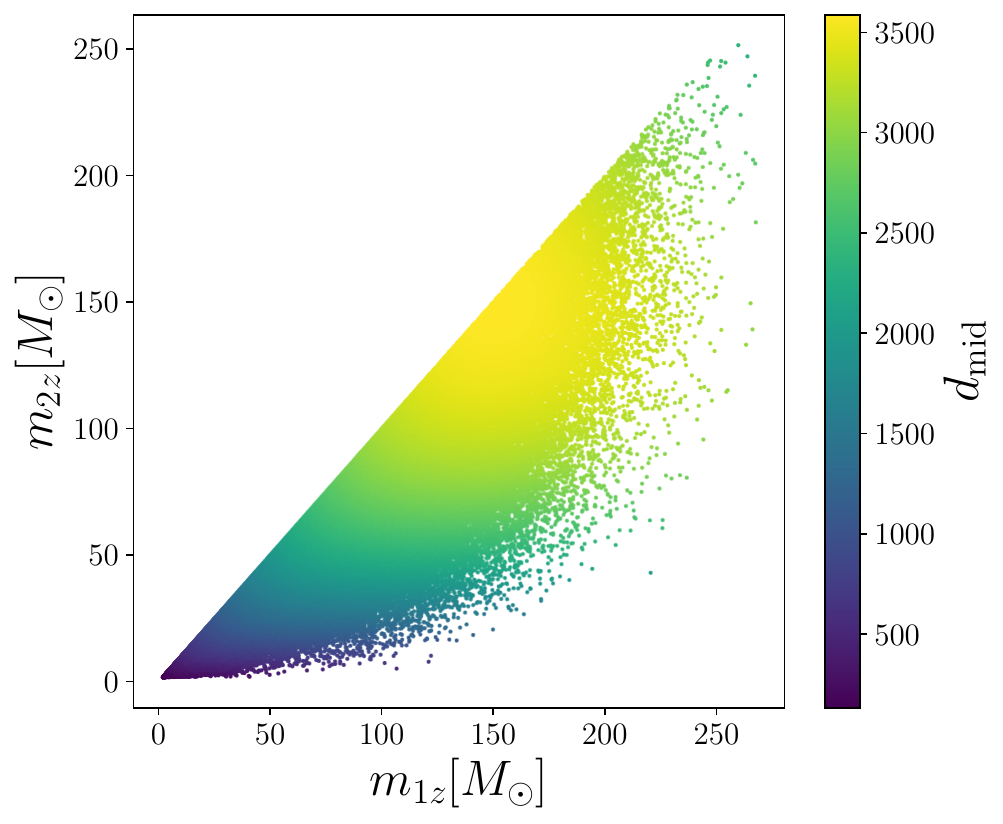}}
	\subfigure[$\dmid$ over redshifted $\mathcal{M}$ and $\eta$.]{ \includegraphics[width=0.49\textwidth]{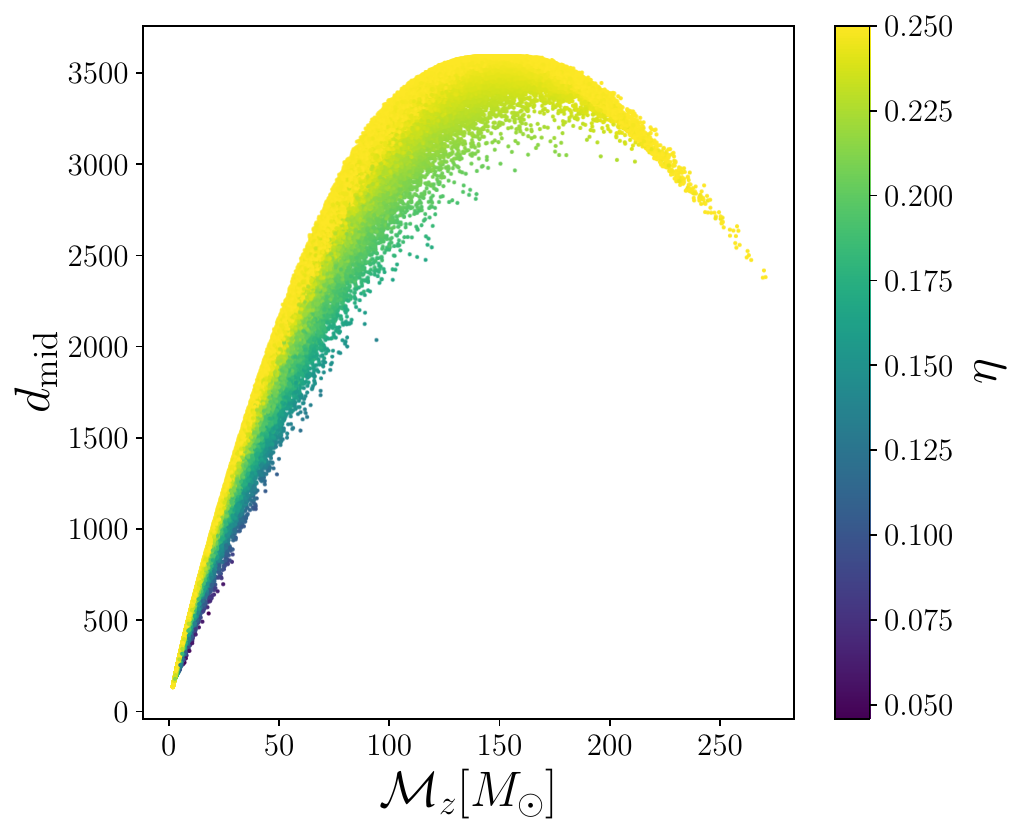}}
	\caption{Sensitive distance $\dmid$ values for the O3 BBH injections obtained from the global fit of $\pdet$, plotted over redshifted masses $m_{1z}, m_{2z}$ (a) and over redshifted chirp mass $\mathcal{M}_z$ and symmetric mass ratio $\eta$ (b).}
	\label{fig:dmid}
\end{figure}

Figure \ref{fig:dmid}(a) shows the values of $\dmid$ obtained from the fit in terms of the redshifted masses, computed for every injection in the GWTC-3 catalog. The dependence on masses presents a local maximum around \td{$\unit[3800]{Mpc}$ for $m_{1z}, m_{2z} \gtrsim \unit[150]{M_\odot}$, which corresponds to redshifted chirp mass $\sim\! \unit[150]{M_\odot}$,} consistent with aLIGO expected sensitivity to compact binary coalescences \cite{aLigo_sensitivity}. For chirp masses below $\sim\! \unit[150]{M_\odot}$ the sensitive distance increases with mass, \td{mainly driven by a higher signal amplitude}.  
However, the merger also shifts towards lower frequencies, eventually leaving the detectors' frequency band, thus the \td{sensitive} distance decreases for chirp masses $\gtrsim \unit[150]{M_\odot}$. \td{Moreover, higher mass signals have fewer cycles visible in band, thus being more readily mimicked by glitches: an effectively higher background for high-mass templates is therefore expected to contribute to decreasing sensitivity towards high masses. 
Figure.~\ref{fig:dmid}(b) shows behaviour consistent with these expectations, and also indicates a trend for higher (lower) sensitivity to equal (unequal)-mass binaries} at a given chirp mass. 

\begin{figure}[htbp]
	\centering
    \subfigure[$\emax$ over total redshifted mass.]
	{\includegraphics[width=0.465\textwidth]{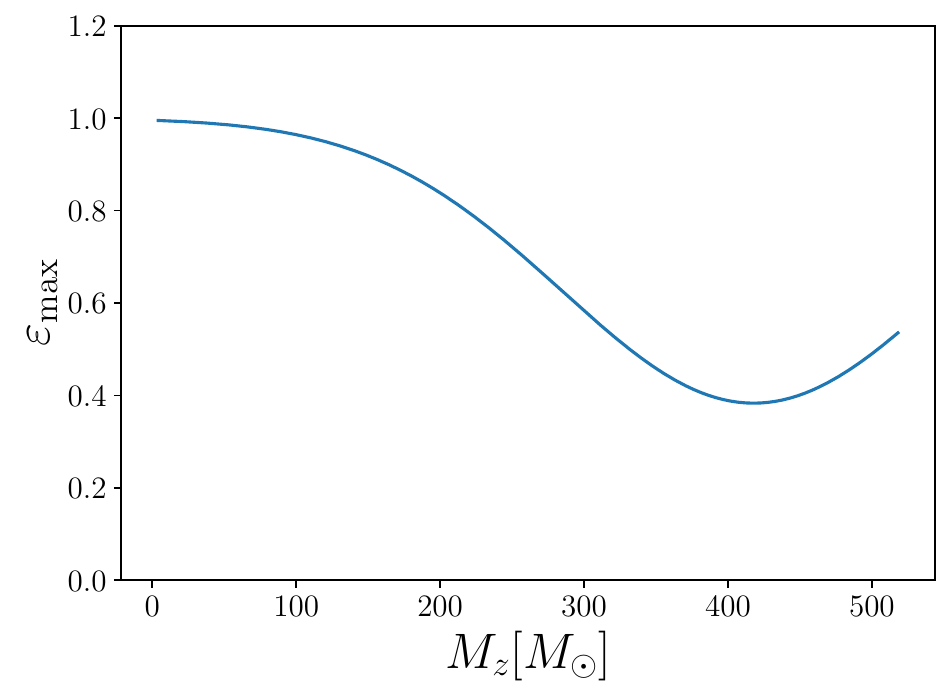}}
    \subfigure[$\pdet$ over distance rescaled to $\dmid$ and over $\emax$.]
    {\includegraphics[width=0.525\textwidth]{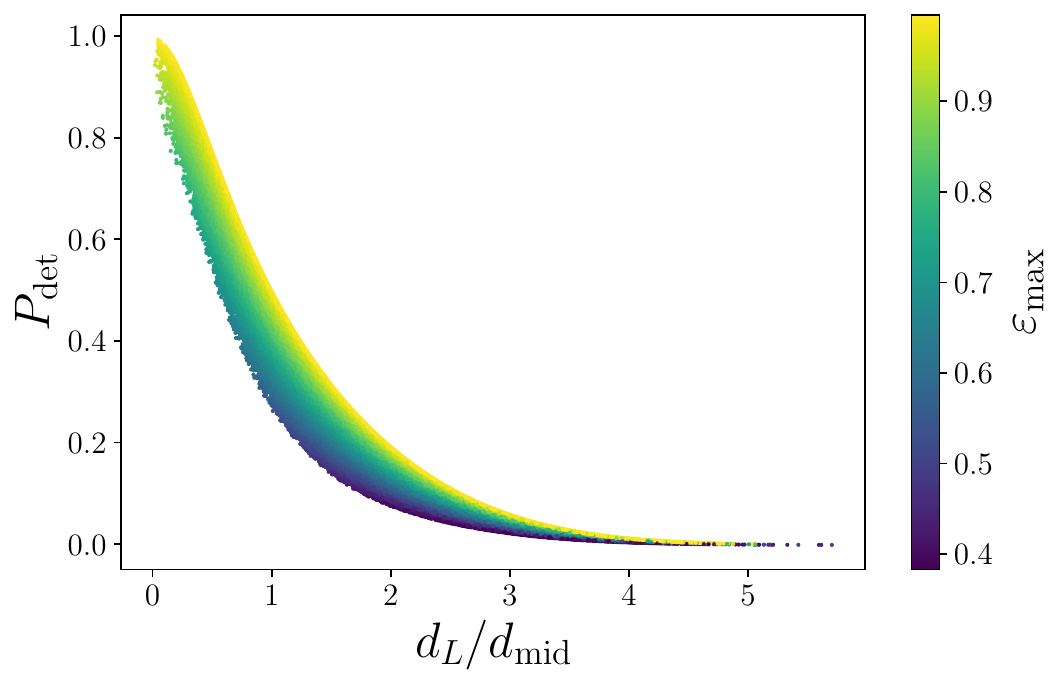}}
	\caption{Mass dependence of the maximum detection probability $\emax$ (a) and probability of detection over distance rescaled to $\dmid$, colored by $\emax$ (b) for O3 BBH injections. \td{The spread of points in (b) indicating a departure from a single universal function is caused by the mass dependence of $\emax$.}}
	\label{fig:emax}
\end{figure}

Maximum detection probability is plotted in Fig.~\ref{fig:emax}(a) in terms of redshifted total mass, presenting an interesting behaviour for higher masses.  \td{The interpretation of the local minimum around $M_z \sim \unit[400]{M_\odot}$ is not yet clear, since various different effects (precessing spins, non-dominant signal modes, higher glitch backgrounds) could lead to high-mass signals not being detected, even at very low distances: we further discuss spin effects in Sec.~\ref{sec:spins}.}  However the number of injections performed at high masses is much smaller, let alone the number detected, making it difficult to obtain precise results in the large mass limit; statistical uncertainties will be discussed in detail in Sec.~\ref{sec:uncertainties}. 

Lastly, Fig.~\ref{fig:emax}(b) shows the probability of detection of every O3 BBH injection using the optimized parameters obtained from the global fit.  Here, we plot distance rescaled to $\dmid$ (i.e.\ removing most of the mass dependence), thus $\pdet$ follows a expected sigmoid shape going up to almost 1 for small $d_L / \dmid$ values and becoming almost zero for very large distances. 




\subsection{Diagnostics: cumulative distributions}

\begin{figure}[tbp]
	\centering
	\subfigure[Cumulative found injection count over $d_L$.]{\includegraphics[width=0.49\textwidth]{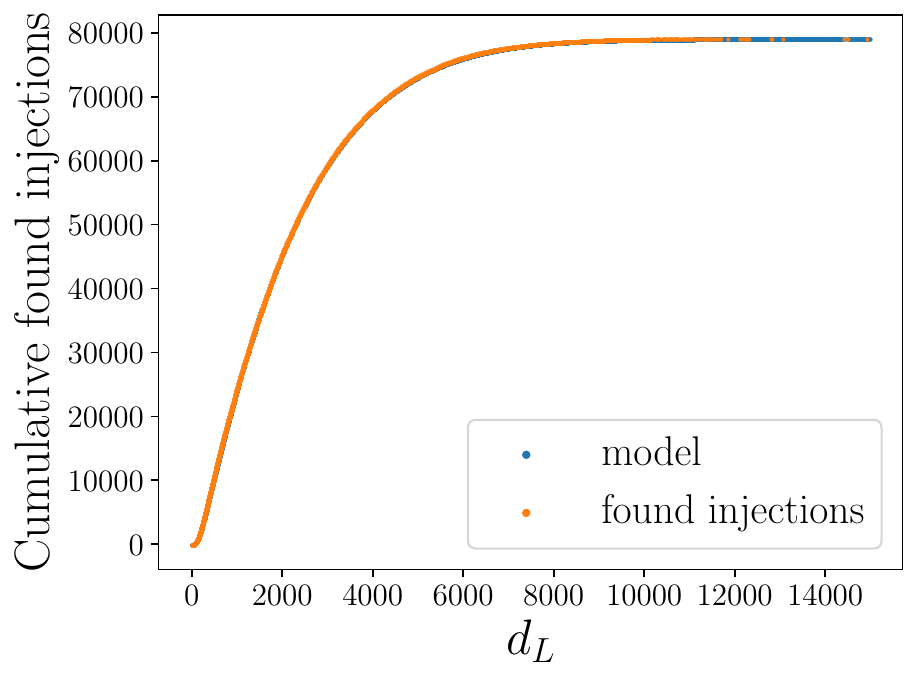}}
	\subfigure[Cumulative found injection count over $\mathcal{M}_z$.]{ \includegraphics[width=0.49\textwidth]{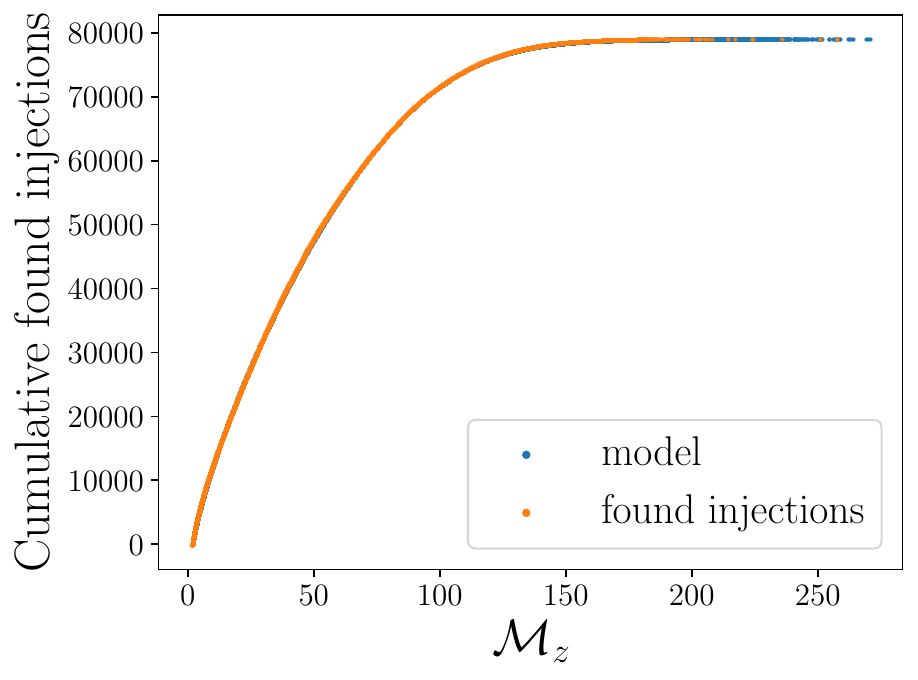}}
	\caption{Cumulative found injections from the GWTC-3 data (orange) and the predicted cdf ('model') from Eq.~(\ref{cdf}) (blue), \td{evaluating the model count at the parameter value of each injection}.}
	\label{fig:cdfs}
\end{figure}

In order to check our choices of functional forms for both $\dmid$ and $\emax$, and also to validate the resulting fit, we compared the cumulative \td{counts and (normalized) distributions (cdfs)} of actual found injections in the GWTC-3 dataset \td{over various parameters of interest} with the \td{cumulative counts or cdfs that are expected from} the inferred selection function.  The expected cdf of found injections is given by 
\begin{equation}
    \mathrm{cdf} (\theta^*) = \frac{\displaystyle 
    \int^{\theta^*}_{-\infty} \pdet (\theta) p(\theta)\, d\theta}{\displaystyle 
    \int^\infty_{-\infty} \pdet (\theta) p(\theta)\, d\theta} \simeq \frac{\displaystyle 
    \sum_{\theta_i \leq \theta^*} \pdet (\theta_i)}{\displaystyle 
    \sum_{\theta_i} \pdet (\theta_i) }\, ,
    \label{cdf}
\end{equation}
\td{i.e.\ the probability of the parameter $\theta$ for a found injection} taking on a value less than or equal to $\theta^*$.  \td{We carried out comparisons} for different variables $\theta$: luminosity distance, symmetric mass ratio, total mass and chirp mass in both the source and the detector frame; for the sake of space, we only illustrate two here. Figure \ref{fig:cdfs} compares the cumulative \td{counts} $\sum \pdet (\theta_i)$ of found injections over luminosity distance and redshifted chirp mass, validating our $\pdet$ fit since the predicted found injections match almost perfectly the actual detected ones. 

Lastly, Kolmogorov-Smirnov (K-S) tests were performed for each of the cdfs \td{to quantify the accuracy of the fit}. Over $d_L$, we obtained a K-S statistic of 0.0026 and a $p$-value of 0.65; similar results for $\mathcal{M}_z$, K-S statistic = 0.0044 and $p$-value = 0.098. Overall, all the K-S statistics values obtained were $0.004$ or less, i.e.\ a maximum deviation $\leq 0.4\%$ of the actual from the predicted distribution, and $p$-values of $0.09$ or larger, i.e.\ no significant discrepancies. Together, the present findings confirm that our model is able to accurately predict the number of found injections, and thus support the choice of the selection function. 
We are, though, unable to constrain or verify the behaviour of $\pdet$ outside the physical range of masses and distances covered by the injection set: in extreme cases (for instance binaries with mass ratio $q$ as small as $\mathcal{O}(0.01)$) our model may yield an unphysical estimate.  However its behaviour may readily be checked within the domain where, for instance, LVK give BBH merger rates estimates\td{~\cite{o3b_pop}}. 

Even though these results show that the fit over the mass and distance degrees of freedom is good, we have ignored any dependence on spins. \td{In fact, comparing cdfs and performing K-S tests over spin variables shows significant discrepancies: this motivates including at least the leading dependence of search sensitivity on component spins, which we describe in the following section.}

\section{Inclusion of leading spin effects}
\label{sec:spins}

Up to this point we are not taking into account spin effects, which would only be reasonable if we assume all black hole spins are small. Further work is needed in order to fit aligned spin dependence to the selection function, as well as distance and masses. The spins of individual black holes within a binary system introduce complexities in the emitted gravitational wave, influencing the amplitude, waveform modulation, and precession of the orbital plane. These spin-induced effects not only contribute to the quality of the observed signals but also impact the accuracy of parameter estimation, effective detection ranges of gravitational wave detectors, and the overall event rates \cite{spin_PE, spin_pipeline, spin_rates}. BBH spins, particularly the effective spin $\chi_\mathrm{eff}$, are essential for extracting precise astrophysical information, providing valuable insights into the population properties and formation channels of binary black hole systems \cite{spin_population, spin_formation_channels}. The effective inspiral spin parameter $\chi_\mathrm{eff}$ is defined as \cite{chieff}
\begin{equation}
    \chi_\mathrm{eff} = \frac{(\vec{s}_1m_1 + \vec{s}_2m_2) \cdot \bm{\hat{L}}}{m_1 + m_2} \equiv
    \frac{m_1 s_{1z} + m_2 s_{2z}}{m_1 + m_2}\,,
    \label{chieff}
\end{equation}
where $\vec{s}_{1,2}$ are the (dimensionless) spins of each black hole and $\hat{L}$ is the direction of the 
orbital angular momentum; this defines the spin projections along the orbital axis $s_{1z}, s_{2z}$. 

\begin{figure}[tbp]
	\centering
	\subfigure[Spin-dependent correction $f_{AS}$ over $\chi_\mathrm{eff}$ and $M_{z}$.]{\includegraphics[width=0.54\textwidth]{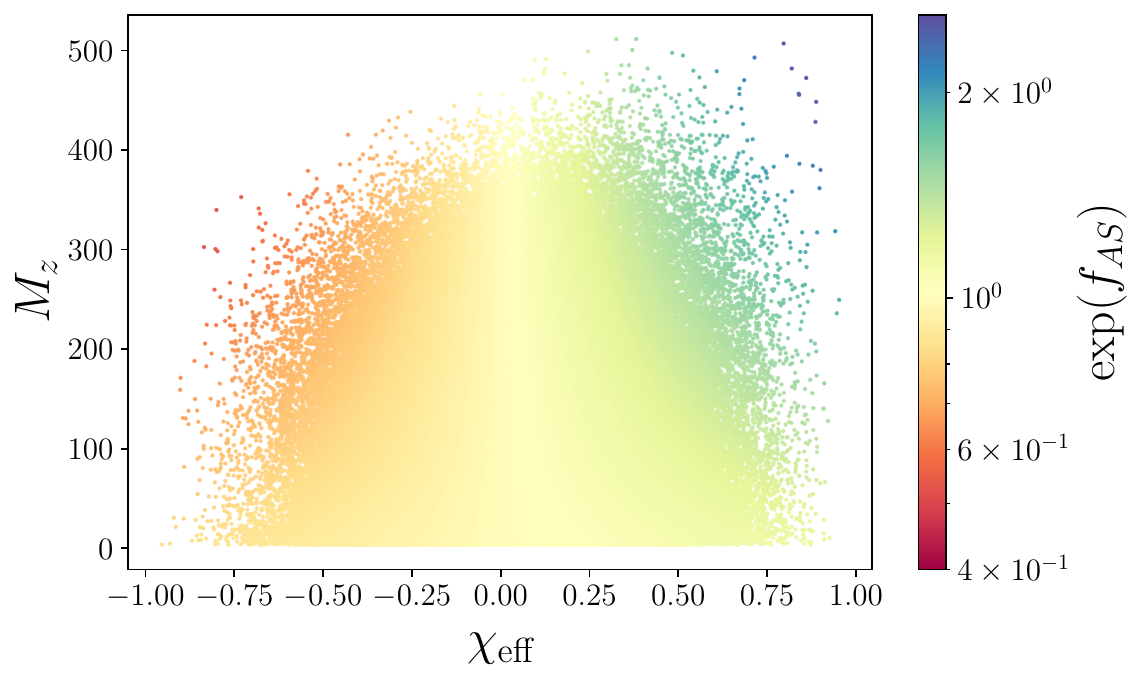}}
	\subfigure[$\emax$ as a function of $M_z$.]{\includegraphics[width=0.44\textwidth]{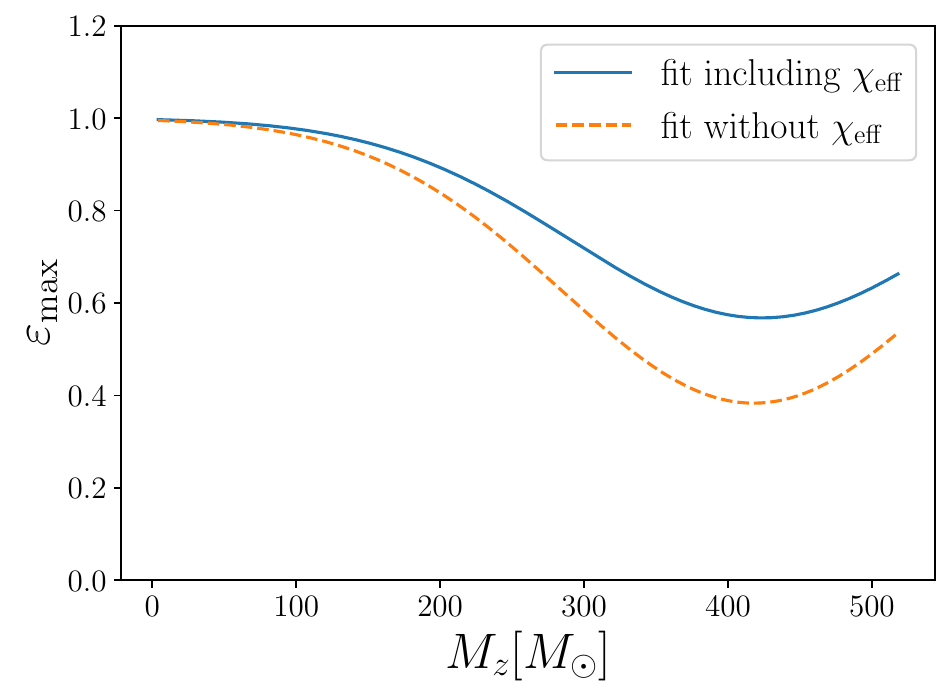}}
	\caption{Search sensitivity dependencies for the fit including aligned spin effects. \td{Panel (a) shows the correction to the sensitive distance, $f_{AS}$, over total redshifted mass and effective spin, for every O3 BBH injection. The comparison between the maximum search efficiency dependence on redshifted total mass from the fit with and without spin correction in shown in (b). }}
	\label{fig:chieff_corr}
\end{figure}

Signals from binaries with $\chi_\mathrm{eff} > 0$ (i.e.\ aligned spins) have a higher probability of detection, since their waveforms have a larger number of cycles and a higher duration in the detector sensitive band~\cite{chieff}; thus $\chi_\mathrm{eff}$ should have a significant impact on the selection function~\cite{chieff_effects,Hoy:2021rfv}. As mentioned previously, we compared cdfs of found injections, as well as histogramming them over various parameters, in order to study the dependence of $\pdet$ on $\chi_\mathrm{eff}$. Significant discrepancies were found with the previous fit function neglecting spins, Eqs.~(\ref{eq:dmid_fd}-\ref{eq:emax_nospin}), 
verifying that the detection probability is an increasing function of $\chi_\mathrm{eff}$, an effect appearing stronger at higher total mass. Taking into account these discrepancies, spin dependence was implemented as follows:\footnote{Here the $c$ subscripts represent powers of $M_z$ and $\chi_\mathrm{eff}$.}
\begin{equation}
   \tilde{d}_\mathrm{mid} = D_0 \, \mathcal{M}_z^{5/6} \, \exp (f_d) \, \exp (f_{AS}) \, , 
   \label{final_dmid}
\end{equation}
\begin{equation}
   f_{AS} = \chi_\mathrm{eff} (c_{01} + c_{11} M_z) \, .
\end{equation}
We thus require to fit the coefficients $c_{01}$, $c_{11}$ simultaneously with other parameters. Some changes to the likelihood are needed to account for the spin dependence $\chi_\mathrm{eff}(m_1, m_2, s_{1z}, s_{2z})$: in place of Eqs.~(\ref{global_likelihood}-\ref{nexp}) we have:
\td{
\begin{equation}
    \ln \mathcal{L} = -\mu + \sum_i \ln \lambda (d_L^i, m_1^i , m_2^i, s_{1z}^i, s_{2z}^i) \, ,
    \label{global_likelihood_spins}
\end{equation}
\begin{equation}
\fl \quad \lambda(d_L,m_1 , m_2, s_{1z}, s_{2z}) = \pdet(d_L, m_1, m_2, \chi_\mathrm{eff}) p(d_L) p(m_1, m_2) p(s_{1z}, s_{2z}) N_{\mathrm{tot}} \, ,
    \label{smalllambda_spins}
\end{equation}
\begin{equation}
    \mu \equiv N_{\exp} = \int\! 
    \lambda(d_L, m_1 , m_2, s_{1z}, s_{2z})\, \mathrm{d}m_1\, \mathrm{d}m_2\, \mathrm{d} s_{1z}\, \mathrm{d} s_{2z}\, \mathrm{d}d_L \, .
    \label{nexp_spins}
\end{equation}
}
\begin{figure}[tbp]
	\centering
	\subfigure[$\dmid$ in terms of $m_{1z}$ and $m_{2z}$.]{\includegraphics[width=0.475\textwidth]{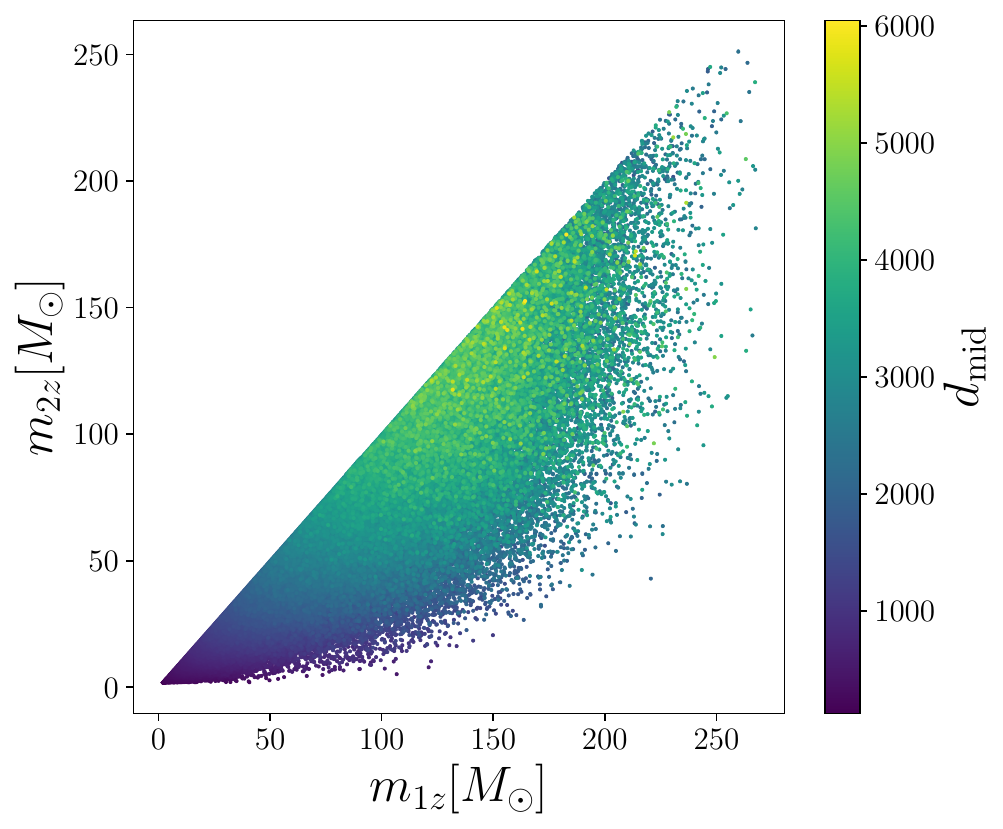}}
	\subfigure[$\dmid$ over redshifted chirp mass and $\chi_\mathrm{eff}$.]{\includegraphics[width=0.49\textwidth]{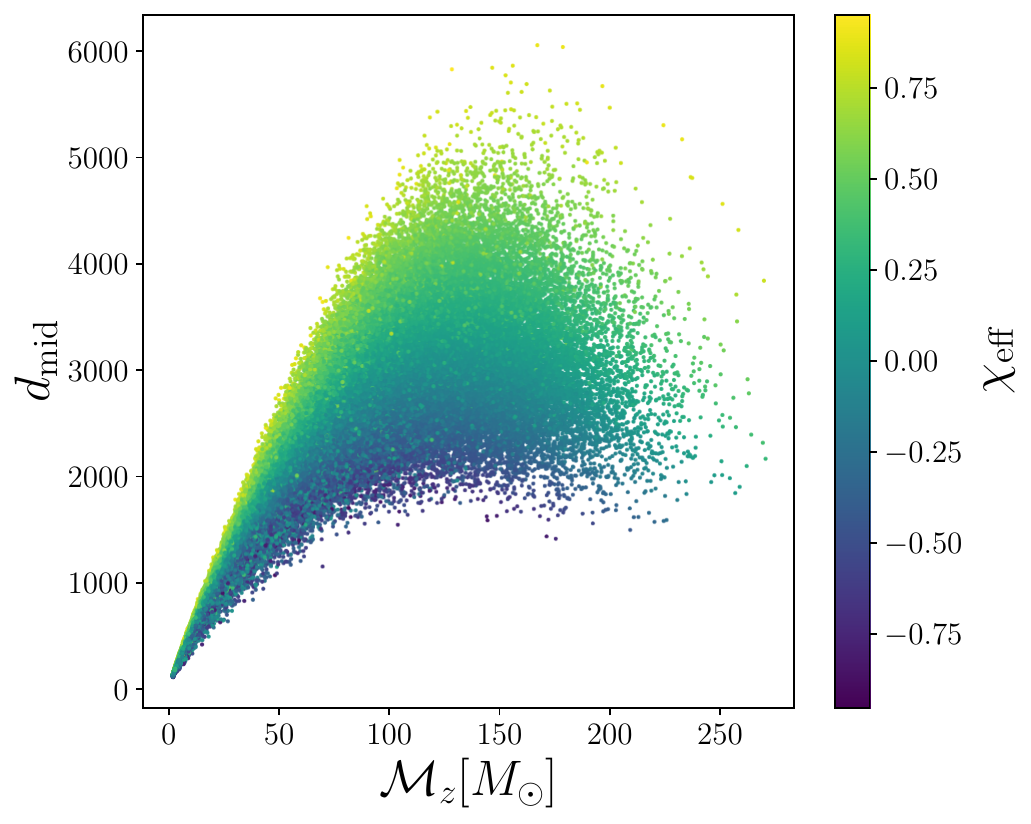}}
	\caption{Sensitive distance $\dmid$ values for the O3 BBH injections from the fit with spin dependence, over redshifted masses (a) and over redshifted chirp mass and effective spin (b).}
	\label{fig:dmid_chieff}
\end{figure}

\td{We find that} the spin-dependent factor $\exp (f_{AS})$, shown in Fig.~\ref{fig:chieff_corr}(a), is close to 1 for small values of $\chi_\mathrm{eff}$ and small masses, but in the high mass limit, extreme values of the effective spin the correction can reduce $\dmid$ by a factor of $\sim 2$ for the most negative values of $\chi_\mathrm{eff}$ and increase it by a factor of $\sim 2.5$ for $\chi_\mathrm{eff} \sim 1$. The spin correction also affected the fit of the maximum detection probability, as evident from Fig.~\ref{fig:chieff_corr}(b). It is plausible that the lower values of $\emax$ from the \td{fit neglecting spins} were driven by the need to fit a mixture of different $\dmid$ values due to different spins at the same component mass point. \td{The remaining apparent trend in $\emax$ is still subject to high uncertainties, explored in Sec.~\ref{sec:uncertainties}.}

Figure~\ref{fig:dmid_chieff}(a) shows the $\dmid$ values in terms of the redshifted masses from the new fit, where we can see that the \td{maximum} sensitive range increases by almost a factor of 2. 
In Fig.~\ref{fig:dmid_chieff}(b) we can see the effect of $\chi_\mathrm{eff}$ in $\dmid$ as a function of chirp mass: basically systems are detectable at higher distances the more aligned the spins are (the closer $\chi_\mathrm{eff}$ is to 1) as expected. By comparison with Fig.~\ref{fig:dmid}(b), at fixed $\mathcal{M}_z$ variations in $\chi_\mathrm{eff}$ appear to have a larger effect on detectability than variations in $\eta$. 

\begin{figure}[tbp]
	\centering
	\subfigure[Cumulative found injections over $\chi_\mathrm{eff}$ without \newline the spin correction.]{\includegraphics[width=0.49\textwidth]{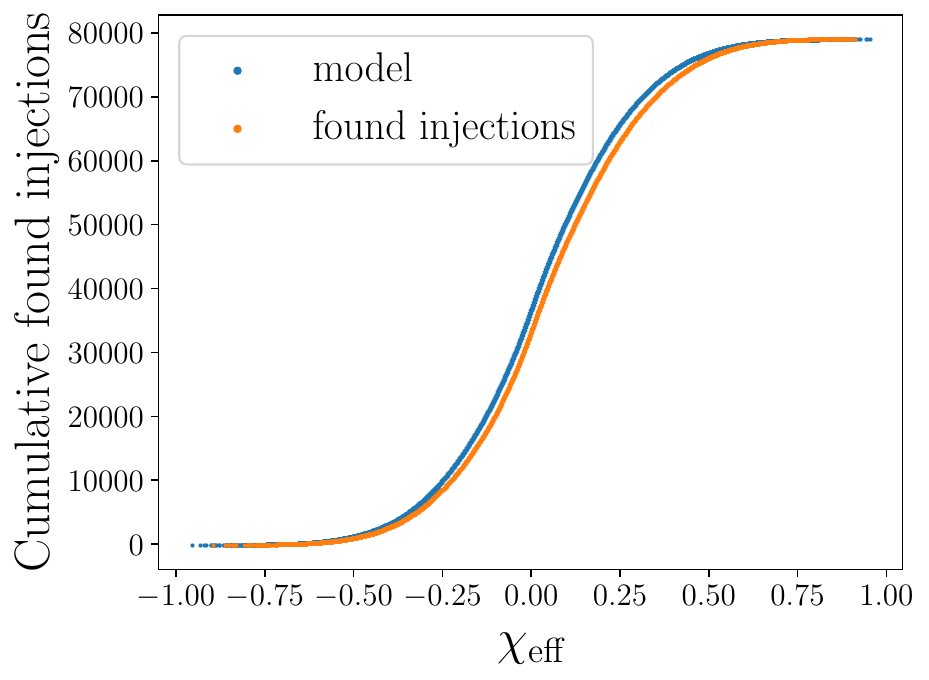}}
	\subfigure[Cumulative found injections over $\chi_\mathrm{eff}$ with \newline the spin correction.]{ \includegraphics[width=0.49\textwidth]{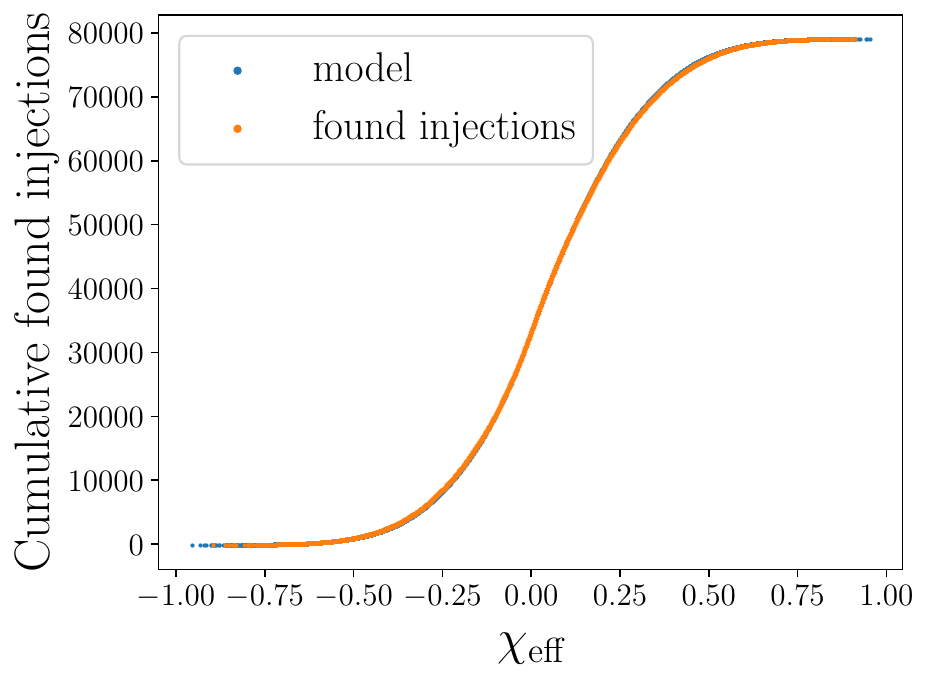}}
	\caption{Comparative of the cumulative found injections over $\chi_\mathrm{eff}$ from the GWTC-3 data (orange) and the predicted cdf from (\ref{cdf}) (blue), without (a) and with (b) spin correction. The K-S statistics and $p$-values for the model without spin correction (a) are $(0.0438, 0)$, and $(0.0039, 0.19)$ for the new model (b).}
	\label{fig:cdfs_chieff}
\end{figure}

Lastly, we perform K-S tests to validate this new model with spin dependence. Figure \ref{fig:cdfs_chieff} shows the cumulative found injections in the GWTC-3 catalog over $\chi_\mathrm{eff}$, along with the cdfs predicted by the selection function, \td{for our fits both without the spin factor (a), and including it (b).  A significant mismatch is visible in the fit without spin dependence}.  In general, the K-S statistic values over different variables obtained with the spin correction were $0.004$ or less again, and $p$-values of $0.10$ or larger, implying that the spin dependence is well fitted.  

\section{Statistical uncertainties in $\pdet$}
\label{sec:uncertainties}

\begin{figure}[tbp]
	\centering
	\subfigure[Uncertainty in the dependence of $\dmid$ on $M_z$, for various fixed $\eta$ values.]{\includegraphics[width=0.47\textwidth]{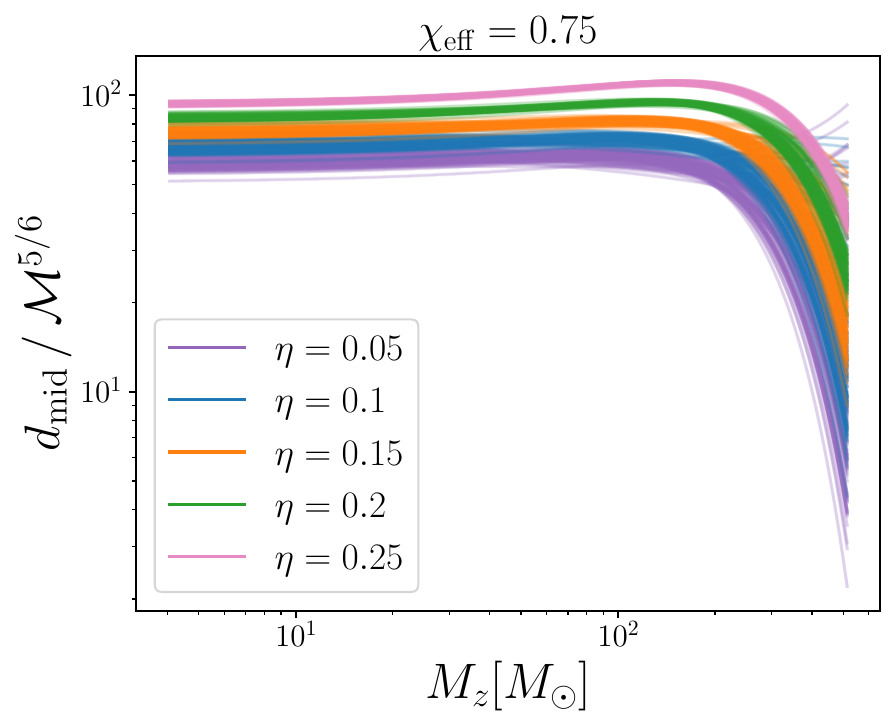}}
	\hspace{0.3cm}
    \subfigure[Uncertainty in the dependence of $\dmid$ on $\chi_\mathrm{eff}$, for various fixed $M_z$ values.]
    {\includegraphics[width=0.47\textwidth]{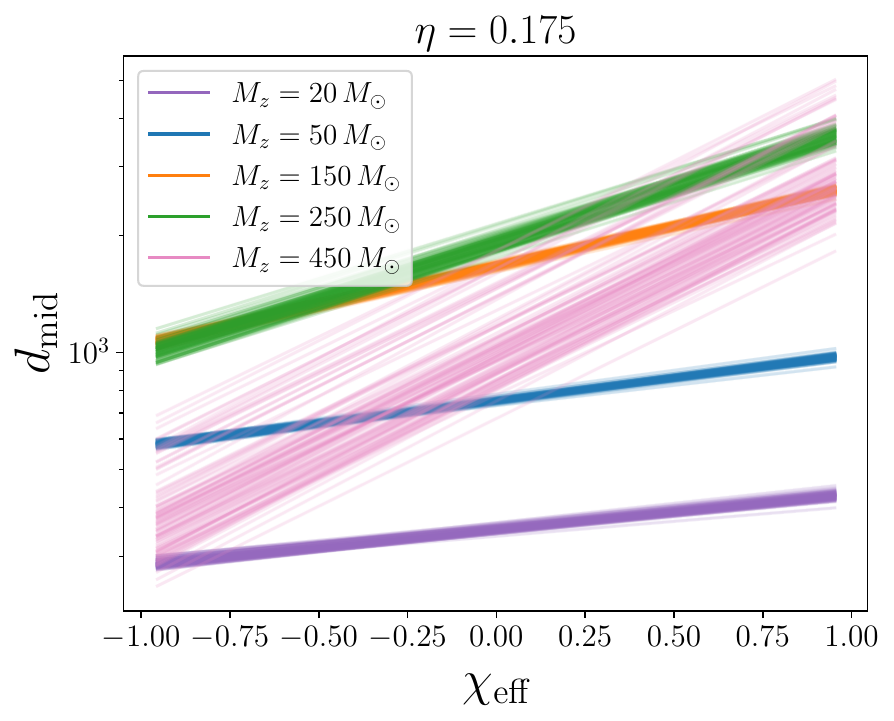}}
    \subfigure[Uncertainty in $\emax(M_z)$.]{\includegraphics[width=0.47\textwidth]{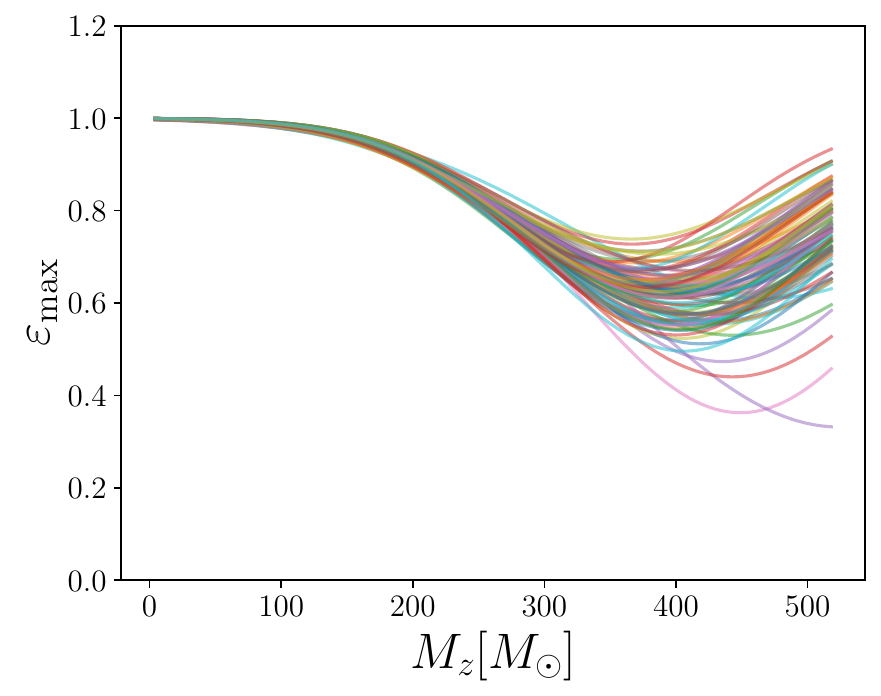}}
	\caption{Uncertainty in $\pdet$ functions quantified via $100$ bootstrap resampled analyses: each plotted line represents a bootstrap iteration.}
	\label{fig:dmid_emax_boots}
\end{figure}

In order to quantify the statistical uncertainties in our estimation of the 
selection function, due to the finite number both of injections performed and, particularly, of found injections, we use a bootstrap resampling method.
We randomly resample the observed data $n$ times, allowing for replacements, and fit $\pdet$ to each bootstrap set of injections, thus obtaining $n$ sets of optimized fit parameters. In addition to estimating the uncertainty on each fit parameter, bootstrap resampling fit allows us to assess how far uncertainties in the parameters are correlated with each other. Figure \ref{fig:dmid_emax_boots} illustrates the resulting uncertainty estimates in $\dmid$ and $\emax$ over the detected total mass and the effective spin, by plotting $n = 100$ bootstrap iterations. 

The results show that the estimated statistical uncertainty is \td{lowest for low masses}, becoming significant in the high mass limit; \td{for a few bootstrap iterations we even see instability in the behaviour at very high total mass (over $\sim\!\unit[200]{M_\odot}$) for unequal mass binaries, which we attribute to a very small number of found injections in this region.  Figure \ref{fig:dmid_emax_boots}(a) additionally shows} that at fixed total mass the uncertainty decreases with increasing $\eta$, due to an increasing density of found injections. \td{Figure \ref{fig:dmid_emax_boots}(b), showing the uncertainty in $\dmid$ as a function of effective spin for different values of the redshifted total mass, again demonstrates that uncertainties grow with $M_z$, but remain near-constant over $\chi_\mathrm{eff}$.  Figure \ref{fig:dmid_emax_boots}(c) shows the uncertainty in the dependence of $\emax$ on $M_z$: while there is a significant decrease from masses $\unit[100]{M_\odot}$ up to $\unit[350-400]{M_\odot}$, the behaviour at very high $M_z$ is subject to large uncertainties, with a slight (not highly significant) trend to increasing $\emax$.}

\td{Most of} these trends are expected, as the number of detected injections in the high (redshifted) mass limit is very low. 
\td{Obtaining useful sensitivity estimates in this regime will require a much higher density of injections here, allowing us to diagnose more clearly what factors may be causing the apparent increase in $\emax$.} It will be important to account for such uncertainties in any population analysis over this mass range; \td{note that the main LVK O4 population analyses did not explore such high masses~\cite{o3b_pop} given the lack of clear detections.}

\section{Discussion}\label{sec:discussion}

In this work we have introduced a simple but accurate, physically motivated approximation of the detection probability $\pdet$ for binary merger signals in LIGO-Virgo-KAGRA search pipelines analyzing real gravitational-wave data, accounting for dependence on binary masses, orbit-aligned spins and luminosity distance.  Observation selection bias, which is quantified via this probability, has a very strong influence on the set of events recovered in any GW catalog, and thus must be taken into account in any kind of population analysis.  The expected search sensitivity to a signal with given masses and spins depends on many diverse factors, including the distance and other extrinsic parameters, but also the detector sensitivities, data quality, and the configuration of search pipelines. These effects are most often accounted for by resampling (importance sampling) of simulated signals, i.e.\ injection sets, but this method is inefficient for general forms of population model, and may incur high uncertainties.  Our proposed fit gives the probability of detection as a continuous (and differentiable) function over mass, spin and distance (or redshift), and thus makes it possible to do population studies using models of arbitrarily complex form without increased uncertainty in the selection function~\cite{Talbot:2023pex}.  The fit may also be used for ``forward modelling'' to directly predict the number and distribution of observed events given the outputs of astrophysical population synthesis calculations. 

Our fit function is motivated by the known dependence of the signal-to-noise ratio, for a binary at fixed distance, on other parameters, mainly the sky location and binary orientation and the chirp mass.  We allow for mass- and spin-dependent corrections to the theoretically expected behaviour: these corrections, as well as an overall scaling (related to the detector network sensitivity), are determined via a maximum-likelihood fit to the O3 injection campaign results.  A surprisingly small number of correction terms is needed to obtain physically reasonable results which accurately predict of the number and distribution found injections, as verified by CDF comparisons and KS tests.  The statistical uncertainty in the fitted $\pdet$ is small, except for the high (redshifted) mass \td{and unequal mass} limits where detected injections are very sparse. 
A limitation of our procedure \td{(common to any empirical estimate of $\pdet$ from search results)} is that we are unable to constrain or verify the behaviour of $\pdet$ outside the physical range of masses, distances and spins covered by the injection set, thus care is required if applying the fit in extreme cases such as the high mass limit.   \td{The limitation would of course be lifted by use of a wider injection parameter range.} 

A version of this fit (without including spin effects) \td{was an essential ingredient for a nonparametric analysis} of the evolution of the BBH population over redshift \cite{Rinaldi:2023bbd}.  While we have concentrated on fitting the results of injections analyzed by search pipelines in real O3 data, the fitting method may also be applied to ``semianalytic'' injections associated with the O1 and O2 runs \cite{o1o2o3_inj_7890398}.  There are, however, significant differences relative to O3 in the method used for identifying these injections as detected -- mainly, thresholding on optimal (expected) network SNR rather than on the search pipeline significance, i.e.\ false alarm rate (FAR) or ranking statistic, as well as the use of constant detector PSDs for each run.  Although the resulting integrated sensitivity over O1 and O2 is consistent with total merger rate estimates, these method differences may produce a bias in the distance dependence of the ``semianalytic'' found injections, compared to what would be found by actual searches of O1 and O2 data with injections.  Therefore, rather than a detailed functional fit of O1-O2 semianalytic injections, we instead recommend keeping the O3 fitted parameters, but scaling the $\dmid$ values by a constant factor for each of the O1 and O2 runs (i.e.\ changing only $D_0$ between runs), such that the total counts of found injections for O1 and O2, and the numbers of actual BBH detections (3 and 10, respectively) are consistently predicted.  

Although our method can already be used for investigations of the existing BBH population, characterized by near-equal masses, relatively low spin magnitudes and generally small or negligible precession effects, there are several possible extensions to its scientific scope.  The next steps will be to include dependence over precessing BH spin components (i.e.\ those in the orbital plane), which are expected to influence search sensitivity, and also investigate the applicability of our fitting method in the mass ranges covered by BNS and NSBH injections. 

Concerning methodological extensions, \td{ways of marginalizing over or otherwise including uncertainties in $\pdet$ are to be investigated: for instance, when performing Bayesian analysis the fit parameters could be sampled over using the associated likelihood.\footnote{We thank Colm Talbot for this suggestion.}}  We are currently treating extrinsic angular parameters as unknowns to be averaged over: the form of our sigmoid function over distance in Eq.~(\ref{pdet}) is primarily determined by how the signal amplitude depends on such nuisance parameters.  We expect that the probability of detection for a binary at a given distance \emph{and with known sky location and orientation} will be related in a simple way to our fitted parametric form via $\dmid$.  Including the explicit angular dependence may facilitate studies of any deviations from isotropy in the merging binary distribution (e.g.~\cite{Stiskalek:2020wbj,Payne:2020pmc}). \td{Analogously, one could consider quantifying the dependence of $\pdet$ on the state of the detector network (rather than simply averaging over time), cf.~\cite{Essick:2020qpo}, which may be of interest for detector characterization.}

\section*{Acknowledgments}\label{sec:acknowledgments}
We are happy to acknowledge helpful discussions with the LVK Rates and Populations group, in particular Colm Talbot and Amanda Farah, and Vaibhav Tiwari for carefully reading an earlier version of this paper.  This work has been supported by the Spanish Agencia Estatal de Investigación
through the grant PRE2022-102569, funded by MCIN/AEI/10.13039/ 501100011033 and the FSE+. 
This work has received financial support from María de Maeztu grant CEX2023-001318-M funded by MICIU/AEI/10.13039/501100011033, and from the Xunta de Galicia (CIGUS Network of Research Centres) and the European Union.  ALM and TD are supported by research grant PID2020-118635GB-I00 from the Spanish Ministerio de Ciencia e Innovaci{\'o}n. 
This material is based upon work supported by NSF’s LIGO Laboratory which is a major facility fully funded by the National Science Foundation.  

\section*{Data availability}

The data that support the findings of this study are openly available at the following URLs: \url{https://github.com/AnaLorenzoMedina/cbc_pdet}, \url{https://zenodo.org/records/7890437}.

\section*{References}
\bibliographystyle{iopart-num}
\bibliography{bibliography}


\clearpage 
\appendix{}

\section{Injection probability density functions}
\label{appendix:pdf}

In this Appendix we summarize the PDFs of the O3 injected BBH signal distribution over the relevant binary parameters~\cite{o3}.  The source-frame mass distribution is parameterized as power laws in the two component masses:
\begin{equation}
    p(m_1, m_2)=p(m_1) p(m_2|m_1) \, ,
    \label{powerlaw}
\end{equation}
with the restriction that $m_2\leq m_1$ and
\begin{eqnarray}
    &p(m_1)\sim m_1^{\alpha}\, , \hspace{0.5cm} m_{1,min}\leq m_1 \leq m_{1,max} \, , \\
    &p(m_2|m_1)\sim m_2^{\beta} \, , \hspace{0.5cm} m_{2,min}\leq m_2 \leq m_1 \, .
\end{eqnarray}
The BBH injection distribution for the GWTC-3 catalog~\cite{KAGRA:2021vkt} uses $\alpha=-2.53$, $\beta=1$, $m_{1,min}=m_{2,min} = \unit[2]{M_\odot}$ and $m_{1,max}=m_{2,max} = \unit[100]{M_\odot}$.  By normalizing (\ref{powerlaw}) to 1, we obtain the explicit mass distribution
\begin{equation}
    p(m_1,m_2)=m_1^{\alpha}\, m_2^{\beta} \,  \frac{\alpha+1}{m_{max}^{\alpha+1}-m_{min}^{\alpha+1}}\, \frac{\beta+1}{m_1^{\beta+1}-m_{min}^{\beta+1}}\, , \hspace{0.5cm} m_2\leq m_1 \, .
    \label{pm1m2}
\end{equation}
This distribution is normalized over the whole space; however, 
in the context of our mass binned fitting (Sec.~\ref{ss:massbin}), we require PDFs normalized over each bin separately.  Since the relevant integrals are straightforward we do not give the normalizations explicitly.  Note that
for bins crossing the equal-mass line ($m_1 = m_2$) precautions are needed, as since half of the bin area has $m_2>m_1$ and thus does not contain injections.  

The redshift distribution used follows the proposal in~\cite{redshiftdist}:
\begin{equation}
    p(z)=\frac{d V_c}{dz} (1+z)^{\kappa-1}  \hspace{0.5cm} z\leq z_{max} \, ,
\end{equation}
where $V_c$ is the contained comoving volume corresponding to a redshift $z$ defined by a flat $\Lambda$CDM cosmology. For BBH injections, $z_{max}=1.9$ and $\kappa=1$. The luminosity distance is related to the cosmological redshift $z$ in flat Lambda-Cold Dark Matter ($\Lambda$CDM) cosmology~\cite{Hogg:1999ad,dl} \dl{via the comoving distance $d_C$ as
\begin{equation}\label{dlformula}
    d_L(z) = (1+z) d_C = (1+z) \frac{c}{H_0} \int_0^z \frac{dt}{\sqrt{\Omega_m (1+t)^3 + (1-\Omega_m)}} \, ,
\end{equation}
where we take $H_0=\unit[67.9]{km/s/Mpc}$, $\Omega_m=0.3065$ and $\Omega_{\Lambda}=1-\Omega_m=0.6935$ (following 
\cite{planck}).  The luminosity distance PDF will be
\begin{equation}
    p(d_L) = \frac{p(z)}{\mathrm{d}(d_L) / \mathrm{d}z} 
    = p(z) \left[ d_C +  \frac{c}{H_0} 
    \frac{(1+z)}{\sqrt{\Omega_m(1+z)^3 + (1-\Omega_m)}} \right] ^{-1} \, .
\end{equation}
}

Finally, the spin distribution is defined over Cartesian spin components for each binary component: the magnitude of each spin is uniformly distributed between zero and the maximum value $s_\mathrm{max}$ and the spin directions are isotropic, thus 
\begin{equation}
    p(s_x, s_y, s_z) = \frac{1}{4 \pi (s_x^2 + s_y^2 + s_z^2)  s_\mathrm{max}} \,\ , \,\, 
        s_x^2 + s_y^2 + s_z^2 \leq s_\mathrm{max}^2 \, .
\end{equation}

\end{document}